\documentclass[prb,twocolumn,showpacs,amsmath,amssymb,superscriptaddress]{revtex4-1}
\usepackage{graphicx}
\usepackage{dcolumn}
\usepackage{bm}
\usepackage{bbm}
\usepackage{ulem}
\usepackage{color}
\usepackage{epstopdf}
\newcommand{\bra}[1]{\left<#1\right|}
\newcommand{\ket}[1]{\left|#1\right>}
\begin{document}

\title{Topological crystalline protection in a photonic system}


\author{Jian-Xiao Zhang}
\author{Mikael C. Rechtsman}
\author{Chao-Xing Liu}
\affiliation{Department of Physics, The Pennsylvania State University, University Park, Pennsylvania 16802-6300, USA}

\date{\today}

\begin{abstract}
Topological crystalline insulators are a class of materials with a bulk energy gap and edge or surface modes, which are protected by crystalline symmetry, at their boundaries. They have been realized in electronic systems: in particular, in SnTe. In this work, we propose a mechanism to realize photonic boundary states topologically protected by crystalline symmetry. We map this one-dimensional system to a two-dimensional lattice model with opposite magnetic fields, as well as opposite Chern numbers in its even and odd mirror parity subspaces, thus corresponding to a topological mirror insulator. Furthermore, we test how sensitive and robust edge modes depend on their mirror parity by performing time dependent evolution simulation of edge modes in a photonic setting with realistic experimental parameters.
\end{abstract}

\maketitle

{\it Introduction -}
Symmetry and topology are two fundamental mathematical tools in the classification of states of matter in condensed matter physics. Recently, intensive research interests have been focused on the role of symmetry in the classification of topological states, ever since the discovery of time reversal invariant topological insulators\cite{qi2010quantum, moore2010birth, qi2011topological, hasan2010colloquium, kane2005z, bernevig2006quantum, moore2007topological, fu2007topological,roy2009topological}, in which time reversal plays an essential role. A large variety of symmetry protected topological states have been identified theoretically for different symmetry classes and dimensions \cite{ryu2010topological,schnyder2008classification}
In contrast, the corresponding material realization in experimentally feasible systems of these new topological states has only been limited in several symmetry classes, mainly for time reversal invariant topological insulators\cite{bernevig2006quantumScience, konig2007quantum, hsieh2009observation, zhang2009nature, chen2009experimental, hsieh2009observationScience, chen2009experimental, hsieh2009tunable, xia2009observation, liu2008quantum}
, the quantum anomalous Hall insulators\cite{chang2013experimental, yu2010quantized}
, topological superconductors\cite{mourik2012signatures, fu2008superconducting, das2012zero, lutchyn2010majorana, sau2010generic, alicea2010majorana}
 and topological mirror insulators\cite{fu2011topological, teo2008surface, dziawa2012topological, hsieh2012topological, xu2012observation, tanaka2012experimental}.
Therefore, searching for new topological systems for symmetry protected topological states is vital for the  development of this field.

Topological crystalline insulator (TCI) phases\cite{fu2011topological}, topological phases that are protected by crystalline symmetry, can exist in a large number of crystal structures with different space group symmetries and a classification of TCIs in different point symmetry groups and space symmetry groups have recently been the focus of research\cite{tanaka2012experimental,  dziawa2012topological, xu2012observation, hsieh2012topological, jadaun2012topological, fang2013theory, dong2015classification, liu2014topological}
. Compared to quantum spin-Hall-type topological insulators, which can only occur for spinful fermions, topological crystalline insulators can exist in both fermionic and bosonic systems. This fact opens up the possibility to realize TCIs in various bosonic systems, including photonic, phononic, magnonic and cold atom systems.  As we describe below, photonic systems in particular have generated a great deal of interest as a probe for topological physics.

The realization by Haldane and Raghu\cite{haldane2008possible, raghu2008analogs} that gyromagnetic photonic crystals could exhibit non-trivial topological invariants opened the door to the new field of Òtopological photonicsÓ\cite{lu2014topological} Ð in which the propagation of photons in a dielectric structure is protected in a similar sense to electrons in a crystal lattice.  The first experimental realization of this phenomenon was made in the group of Solja\v{c}i\'c\cite{wang2008reflection, wang2009observation} for the microwave regime.  However, scaling the wavelength down to the optical regime (in order to realize topological states in optical devices) was not possible using this mechanism due to weak magnetic response in that frequency regime.  Other mechanisms were proposed\cite{umucalilar2011artificial,hafezi2011robust,fang2012realizing,khanikaev2013photonic}, and finally experimental demonstrations were made\cite{rechtsman2013photonic} in a system based on evanescently-coupled helical waveguides; as well as in two-dimensional coupled ring resonators\cite{hafezi2013imaging}.  While photonic topological protection is conceptually similar to that of electronic topological protection (after all, this phenomenon comes down to non-interacting wave dynamics), photonics offers unique advantages and potential applications.  To name a few examples, photonic systems can be designed directly by fabrication (allowing any desired lattice structure to be realized); experiments can be carried out at room temperature, meaning any emergent devices can be brought to application more realistically than those that require very low temperatures; and the robustness associated with topological protection could be of use in an array of devices that rely on the flow of light (e.g., sensors, optical interconnects, electrooptic modulators, isolators, among others).

Very recently, a prediction was made that topological photonic crystals with surface states protected by the glide symmetry could be realized for microwave photons\cite{lu2015three} in a macroscopic ferrimagnetic structure.  In this work, we propose the optical realization of crystalline symmetry protected boundary modes in photonic crystals phase using an entirely distinct mechanism from the microwave.  The systems is quasi-one-dimensional, where an auxiliary parameter, $\phi$, of which the Hamiltonian is a function, is used in place of a second spatial dimension.  This is reminiscent of Ref. \onlinecite{kraus2012topological}, in which a family of one-dimensional systems (defined by a pump parameter) were used to realize a photonic topological edge state and pump that are mapped to the integer quantum Hall effect.  We note that another work \cite{wu2015} has predicted an analogue of a quantum spin Hall system that requires $C_6$ crystalline symmetry; however the edge always breaks that symmetry and so the topological edge states are not protected (and are gapped).  The realization of crystalline symmetry protected topological states in the optical regime allows for a novel paradigm in exploring topological crystalline phases, opening up the possibility to explore the relationship between topological photonic crystals and nonlinear/interacting bosonic systems and quantum optical effects such as multi-photon quantum walks.

\begin{figure}
\centering
\includegraphics[width=0.5\textwidth]{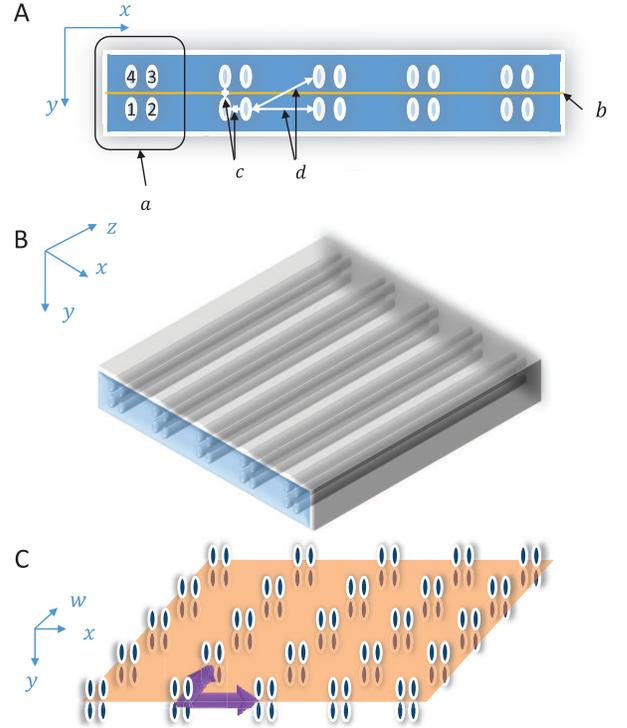}
\caption{(A) Schematics for the quasi-1D chain model. There are four confining potentials, denoted by $1,\dots,4$, in one unit cell, labeled by ``$a$''. Four sites in one unit cell are symmetric with respect to the mirror line ``$b$''. The intra-unit-cell hopping and inter-unit-cell hopping are labeled by ``$c$'' and ``$d$''. (B) Schematics for the 3D visualization of the waveguide chain. The light propagates along $z$ direction. (C) Schematics of the quasi-2D lattice model which can be mapped from the 1D model. The mirror plane $y=0$ (orange) is parallel with the $x-w$ plane. The complex hopping terms along $x$ and $w$ directions between unit cells are symbolized by purple arrows. $w$ is conjugate with the parameter $\theta$. (See Hamiltonian Eq.\,\ref{eq_2d_Hamiltonian}).}
\label{fig_model}
\end{figure}

%
%
%
{\it Tight-binding model - }
We start from a simple tight-binding model to illustrate our main idea and then simulate the system in a more realistic situation. We consider a quasi-one-dimensional (1D) chain along the $x$ direction with each unit cell consisting of four sites, denoted by $\alpha = 1,\cdots,4$, as marked by ``$a$'' in Fig.\,\ref{fig_model}A. The positions of four sites are chosen to preserve $y$-directional mirror symmetry with respect to the line denoted by ``$b$'' in Fig.\,\ref{fig_model}A. Bound states can be induced by lowering on-site energy, forming a potential well on each site. We denote these states as $\ket{s_\alpha}$, and the corresponding creation and annihilation operators as $c^\dag_\alpha$ and $c_\alpha$, respectively. Thus, the tight-binding Hamiltonian for this system is given by

\begin{equation}\label{eq_Ham0}
	H=\sum_{\alpha\beta,m} \left[ c_{m}^{\alpha\dag} \hat V^{\alpha\beta}_{m} c^{\beta}_{m}+\left( c_{m}^{\alpha\dag}  \hat T^{\alpha\beta}_{m} c^{\beta}_{m+1}+\text{h.c.}\right) \right]
\end{equation}
where $\alpha,\beta=1,\cdots,4$ and $m=1,2,\cdots,M$ denoting the unit cell index. $\hat V_m$ describes the on-site energy and the hopping between sites within the $m$-th unit cell. $\hat T_m$ is for the hopping between two nearest neighbor unit cells $m$ and $m+1$. In the  $\{\ket{s_\alpha}\}$ basis, the Hamiltonians $\hat V_m$ and $\hat T_m$ are given by
\begin{equation}\label{eq_Matrix_form_Hamiltonian}
\hat V_m = \left( \begin{array}{cccc}
V_m			&\gamma_x	&0			&\gamma_y\\
\gamma_x	&V_m		&\gamma_y	&0		\\
0			&\gamma_y	&V_m		&\gamma_x\\
\gamma_y	&0			&\gamma_x	&V_m
\end{array} \right),
\hat T_m = \left( \begin{array}{cccc}
0	&t	&t'	&0\\
0	&0	&0	&0\\
0	&0	&0	&0\\
0	&t'	&t	&0
\end{array} \right),
\end{equation}
where $V_m(\phi)$ is the on-site energy of each site in the $m$-th unit cell, chosen to be identical. $\gamma_i$ denotes the hopping strength along the $i=x,y$ direction and $t, t'$ are hopping parameters between two unit cells, with $\gamma_i, t, t' < 0$. Only vertical and horizonal nearest-neighbor(NN) intra-site hoppings, denoted by ``$c$'' in Fig.\,\ref{fig_model}A, as well as horizonal and diagonal NN inter-site hoppings, denoted by ``$d$'' in Fig.\,\ref{fig_model}A, are considered.

\begin{figure}
\centering
\includegraphics[width=0.4\textwidth]{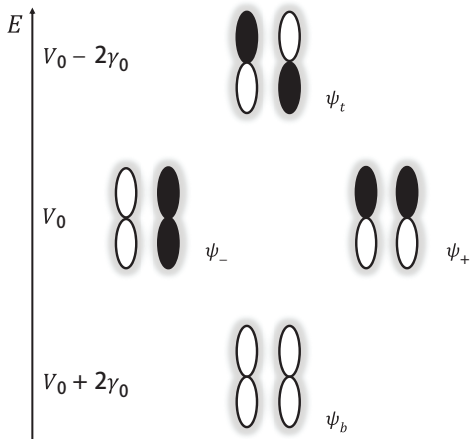}
\caption{ The eigen energies and eigen wavefunctions for the Hamiltonian in one unit-cell with four sites. White and black colors indicate the sign for four different eigen wavefunctions (real).}
\label{fig_eigenstates}
\end{figure}

The above Hamiltonian preserves the $y$-direction mirror operator $\hat{M}_y$, given by
\begin{equation}\label{eq_parity_my}
\hat{M}_y = \left( \begin{array}{cccc}
0	&0	&0	&1\\
0	&0	&1	&0\\
0	&1	&0	&0\\
1	&0	&0	&0
\end{array} \right)
\end{equation}
which interchanges the sites $1$ and $4$ ($2$ and $3$). We set $V_m(\phi)=V_0+\eta_m(\phi)$, $\gamma_x=\gamma_{x0} $, $\gamma_y(\phi)=\gamma_{y0} + \delta_m(\phi)$. Let us neglect $t$, $t'$, $\eta_m$ and $\delta_m$, and set $\gamma_{x0}=\gamma_{y0}=\gamma_0$ for the moment. In this limit, the Hamiltonian $\hat V_m$ in (Eq.\,\ref{eq_Matrix_form_Hamiltonian}) can be diagonalized and the eigen-energies of four eigenstates are
shown in Fig.\,\ref{fig_eigenstates}. We find that two degenerate states $\ket{\psi_+}$ and $\ket{\psi_-}$ with eigenenergy $V_0$ are well separated from other two states $\ket{\psi_{t}}$ and $\ket{\psi_{b}}$ with energies $V_0\mp2\gamma_0$, where we use $+$ ($-$) to represent odd (even) mirror parity of $\hat{M}_y$. 

\begin{figure*}
\centering
\includegraphics[width=1\textwidth]{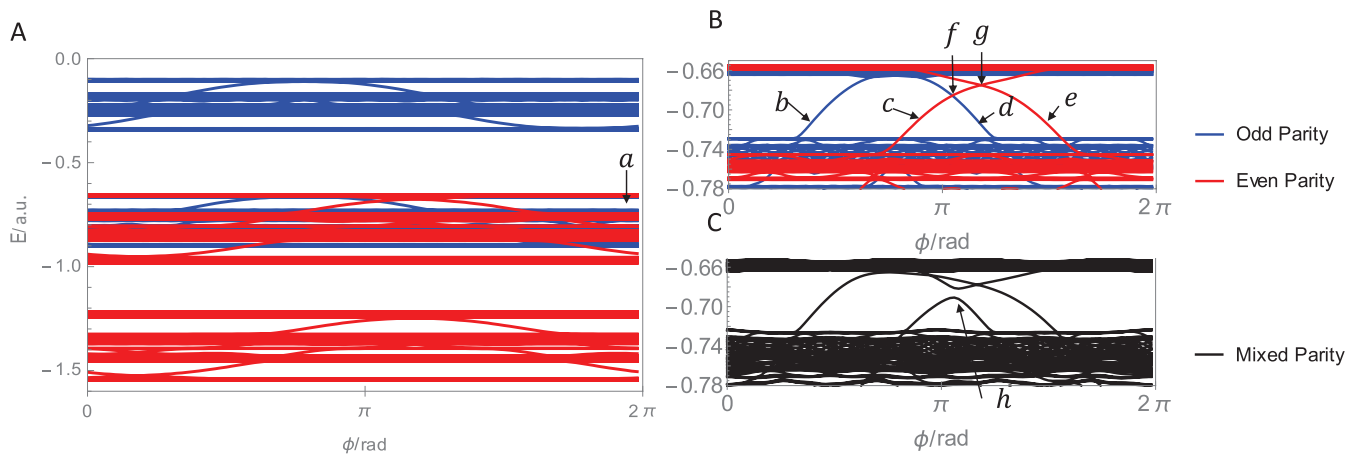}
\caption{(A) Energy spectrum for a finite chain ($M=103, m_0=9	$) of our tight-binding model. Blue and red colors represent odd and even mirror parities of eigen wavefunctions, respectively. There are four bands in energy spectrum. The top and bottom bands come from the states $\ket{\psi_t}$ and $\ket{\psi_b}$, while two bands in the energy range $[-1,-0.6]$ originate from the states $\ket{\psi_\pm}$. A mini-gap, as indicated by ``$a$'', exist within the $\ket{\psi_\pm}$ bands. (B) Zoomed-in graph of energy spectrum around the energy range $[-0.78,-0.66]$. Boundary modes $\ket{\psi_{L+}}$, $\ket{\psi_{R-}}$, $\ket{\psi_{R+}}$, and $\ket{\psi_{L-}}$, as indicated by ``$b$'', ``$c$'', ``$d$'' and ``$e$'', are found within the mini-gap ``$a$''. ``$f$'' (``$g$'') denotes the crossing between ``$c$'' and ``$d$'' (``$c$'' and ``$e$'').
The light red and light blue shading indicate the mini-gap region ``$a$'' of two corresponding parities.
(C) A gap opening for the crossing ``$f$'' can be induced by a mirror-symmetry-breaking term $\Delta H$, leading to an anti-crossing ``$h$''.}
\label{fig_spectrum}
\end{figure*}

Thus, in the limit $ \eta_m, \delta_m, t, t' \ll V_0,\gamma_0$,  we can focus on two degenerate eigenstates $\ket{\psi_+}$ and $\ket{\psi_-}$, and project the total Hamiltonian onto the subspace spanned by these two states. The eigenstates of two degenerate states are given by
$\ket{\psi_+} = \frac{1}{2}(+\ket{s_1}+\ket{s_2}-\ket{s_3}-\ket{s_4})$,
$\ket{\psi_-} = \frac{1}{2}(+\ket{s_1}-\ket{s_2}-\ket{s_3}+\ket{s_4})$,
and let us denote $d_{\pm,m}$ and $d^{\dag}_{\pm,m}$ to be the annihilation and creation operators for the states $\ket{\psi_{\pm}}$ on the $m$-th unit cell, respectively. The effective Hamiltonian after projection is given by
\begin{eqnarray}\label{eq_AAM}
	H=\sum_{\pm,m} t_{\pm,m} (d_{\pm,m}^\dag d_{\pm,m+1}+\text{h.c.})+V_{\pm,m} d_{\pm,m}^\dag d_{\pm,m}
\end{eqnarray}
where the summation $\pm$ is over the parity subspaces. The parameters $t_{\pm,m}$ and $V_{\pm,m}$ are related to the original parameters $t$, $t'$, $\delta_m$ and $V_m$ by $V_{\pm,m}=V_0 + \eta_m \mp \delta_m$ and $t_{\pm} = (-t' \pm t)/2$. If choose $\eta_m = \lambda \cos \phi \cos (2\pi b (m + m_0))$, $\delta_m = \lambda \sin \phi \sin (2\pi b (m+m_0))$,
we find $V_\pm=V_0+\lambda \cos(\phi \pm 2\pi b (m+m_0))$ correspondingly. As a result, the Hamiltonian (Eq.\,\ref{eq_AAM}) is nothing but two copies of Aubry-Andre-Harper (AAH) Hamiltonian \cite{aubry1980analyticity,ganeshan2013topological,kraus2012topological} with effective flux strength $b$ and $-b$ in the odd and even subspaces, respectively. The parameter $\phi$ is also known as the pump parameter. Here $m_0$ shifts the index of the first unit cell, introducing a degree of freedom which will be discussed later. As a consequence of opposite mirror parities under $\hat{M}_y$, these two copies of Hamiltonian are decoupled from each other once mirror symmetry $\hat{M}_y$ is preserved. It has been shown that by correctly choosing parameters, boundary modes can exist at the end of a finite chain described by the AAH model, within fractal sets of band gaps\cite{hofstadter1976energy}. Thus, we perform a direct calculation of energy spectrum of a finite chain for the model (Eq.\,\ref{eq_Ham0}).
Here for demonstration, we choose the parameters to be $\gamma_0 = -0.3, \lambda = 0.1, t=-0.11, t'=-0.033$, $V_0 = -0.8$, $M = 103$, $m_0 = 9$ and $b=\sqrt{5}$. An extra offset between $\gamma_x$ and $\gamma_y$ is set by taking $\gamma_{x0} = \gamma_0 + 0.025$ and $\gamma_{y0} = \gamma_0$. This offset shifts the relative energy level for odd and even bands, in order to emphasize the crossings of edge states inside the mini-gap described later. The four energy levels in one unit cell $\ket{\psi_+}$, $\ket{\psi_-}$, $\ket{\psi_{t}}$ and $\ket{\psi_{b}}$ expand into four energy bands as a function of the parameter $\phi$, as shown in Fig.\,\ref{fig_spectrum}A. The $\ket{\psi_+}$ and $\ket{\psi_-}$ bands overlap with each other in the energy regime $[V_0 -\lambda,V_0+\lambda]$ to the lowest order approximation, which is about $[-1.0,-0.6]$ for the parameters listed above. Each band is split into several sub-bands with mini-gap. We focus on the mini-gap in the energy range [$-0.73, -0.67$] denoted as a in Fig.\,\ref{fig_spectrum}A. There are other mini-gaps, for example in the range of [$-0.95, -0.90$] and [$-1.32, -1.25$]. Here we show the zoomed-in spectrum in Fig.\,\ref{fig_spectrum}B of the mini-gap region ``$a$''. We find four energy levels in the mini-gap, denoted by $\ket{\psi_{L+}}$, $\ket{\psi_{R-}}$, $\ket{\psi_{R+}}$, and $\ket{\psi_{L-}}$, or ``$b$'', ``$c$'', ``$d$'' and ``$e$'' in Fig.\,\ref{fig_spectrum}B, respectively, where L and R denote the position where the wave function of the state is localized in the chain, and $\pm$ indicates the parity of the boundary modes, shown by blue (odd) and red (even) in Fig.\,\ref{fig_spectrum}B. A more realistic calculation of the wave function in this system will be given in the next section, based on simulating the full continuum problem. The crossing between the boundary modes $\ket{\psi_{R-}}$ and $\ket{\psi_{R+}}$, marked by ``$f$'' in Fig.\,\ref{fig_spectrum}B, is topologically protected due to opposite mirror parities between them, while the crossing between $\ket{\psi_{L-}}$ and $\ket{\psi_{R-}}$, marked by ``$g$''
in Fig.\,\ref{fig_spectrum}B, is gapless because these two states are located at opposite boundaries of the chain. As long as the chain size is large compared with the penetration length of boundary mode, which is around twice the lattice constant for the parameters listed above, the overlapping between wavefunctions at opposite boundaries is negligible.

To test the symmetry protection by adding a mirror-symmetry breaking term $\Delta H$ onto $\hat V_m$, where
\begin{equation}\label{eq_symmetry_breaking_H}
\Delta H = \left( \begin{array}{cccc}
0	&\Delta	&0	&0\\
\Delta	&0	&0	&0\\
0	&0	&0	&-\Delta\\
0	&0	&-\Delta	&0
\end{array} \right)
\end{equation}
and $\Delta = 0.04$, we find a gap opening between two edge modes $\psi_{R-}$(``$c$'') and $\psi_{R+}$(``$d$''), as marked by ``$h$'' in Fig.\,\ref{fig_spectrum}C.
Thus, we conclude that these boundary modes are stable only when mirror symmetry is present in the system (Eq.\,\ref{eq_Ham0}).


We would like to emphasize that although our model is written in one dimension, it can be mapped to a two-dimensional (2D) lattice model with complex hopping terms,  corresponding to a 2D topological mirror insulator.  This is a similar mapping as that made in Ref.\,\onlinecite{kraus2012topological}. We may introduce a fictitious dimension $w$ and extend our 1D tight-binding model to the $x-w$ plane, forming a 2D square lattice as shown in Fig.\,\ref{fig_model}C. Similar to the unit cell defined in the 1D model, each unit cell at the integer coordinates $(x,w)$ consists of 4 sites, each having one state described by the creation operators $c^{\alpha\dag}_{x, w}$ where $\alpha = 1,\dots,4$. The 2D Hamitlonian can be written as
\begin{eqnarray}\label{eq_2d_Hamiltonian}
H &=& \sum_{xw,\alpha\beta} \left[\left( c^{\alpha\dag}_{x, w+1}  \hat {T}^{\alpha\beta}_x c^\beta_{x, w} + c^{\alpha\dag}_{x,w+1} \hat {T}^{\alpha\beta}_w c^{\beta}_{x,w} \right) + \text{h.c.}\right] \nonumber\\
&+& \sum_{xw,\alpha\beta }\hat {U}_{\alpha\beta } c^{\alpha\dag}_{x,w} c^\beta_{x,w},
\end{eqnarray}
where $\hat {T}_{x(w)}$ is the hopping matrix between sites labeled by $\alpha,\beta=1,2,3,4$, in the adjacent unit cells along the $x$ ($w$) direction, and $\hat{U}$ is the hopping matrix among sites in one unit cell. The detailed forms of $\hat {T}_{x(w)}$ and $\hat{U}$ are given in the Appendix.

This Hamiltonian is also invariant under the mirror operation $\hat{M}_y$ (Eq.\,\ref{eq_parity_my}) about the $x$-axis. We define the basis creation operators $d^{\dag}_{\pm}$ in the same way as in Eq.\,\ref{eq_AAM}. where $\pm$ are defined for the subspaces with odd and even mirror parities as before. In this new basis, the Hamiltonian takes the form
\begin{eqnarray}
&H = \sum_{\pm,xw}\left[ V_0 d^{\dag}_{\pm,xw} d_{\pm,xw} + \left( (-t' \pm t) d^{\dag}_{\pm,x+1,w}d_{\pm,x,w} \right.\right.\nonumber\\
&\left.\left.+ \frac{\lambda}{2} e^{\pm i\theta} d^{\dag}_{\pm,x,w+1} d_{\pm,xw} + \text{h.c.} \right) \right],
\end{eqnarray}
where the parameter $\theta = 2\pi b x$ represents the phase shift during the hopping along the $w$ direction. We notice that the form of $\theta$ corresponds to the Landau gauge for a magnetic field with the flux $b$ in one unit cell. Thus, a Chern number can be defined in each mirror subspace for a non-zero $b$, as described in the appendix. The opposite signs for the phase shift in the mirror even and odd subspaces indicate that the total Chern number is canceled for the whole system. However, a mirror Chern number, as defined in Appendix, can characterize non-trivial topological property of our 2D system\cite{hsieh2012topological,kraus2012topological}. Therefore, this mapping suggests that our tight-binding model provides a realization of 2D topological mirror insulators.

{\it Numerical Simulations for photonic systems - }
We have now established topological mirror insulator phases in our simple 1D AAH type of tight-binding model and its relation to 2D topological mirror insulators is also illustrated above. The realization of this model in a photonic system requires more sophisticated numerical simulation, which includes continuum degrees of freedom and are based in Maxwells' equations. Next, we describe the detailed experimental setup and perform a numerical simulation of photonic lattice systems.

To realize the tight-binding model in a realistic photonic system, we consider an array of evanescently coupled elliptical waveguides in fused silica glass. The difference in refractive index inside and outside the waveguide is utilized to confine the light in the $x-y$ plane and serve as a potential well, as shown in Fig.\,\ref{fig_model}B. We follow the standard paraxial approximation for this type of waveguides, in which the Maxwells' equation of light can be simplified to a Schr\"odinger-type equation, namely:
 \begin{eqnarray}\label{eq_schrodinger}
i \partial_z \psi(x,y;z) =  &&-\frac{1}{2k_0}(\partial _x^2+\partial_y^2) \psi(x,y;z)\nonumber\\
&&-\frac{k_0\Delta n(x,y;z)}{n_0}\psi(x,y;z)
 \end{eqnarray}
 where $\psi$ is the envelope function of electric field, $k_0$ is the wavenumber of ambient light in the medium, $n_0$ and $\Delta n$ are respectively the background and the waveguide deviation from background refraction index. In comparison with the Schr\"odinger equation, one can see that the $z$ direction can be regarded as time, therefore the diffraction of light through the waveguides is equivalent to the time evolution of a particle in a potential determined by $-\Delta n(x,y;z)$. This equation has been utilized to describe other topological phases in photonic lattices\cite{rechtsman2013photonic}, as well as a wide variety of effects in linear and nonlinear optics\cite{yariv1997optical}, including (but not limited to) the prediction\cite{christodoulides1988discrete} and observation \cite{eisenberg1998discrete,fleischer2003observation} of lattice solitons, stable photorefractive solitons; Anderson localization in optics\cite{schwartz2007transport}; conical diffraction\cite{peleg2007conical}; and optical pseudomagnetism\cite{rechtsman2013strain}.  The waveguides used in this work can potentially be fabricated using the femtosecond direct laser writing technique; and are inspired by those described in Ref.\,\onlinecite{szameit2010discrete}.

\begin{figure}
\centering
\includegraphics[width=0.5\textwidth]{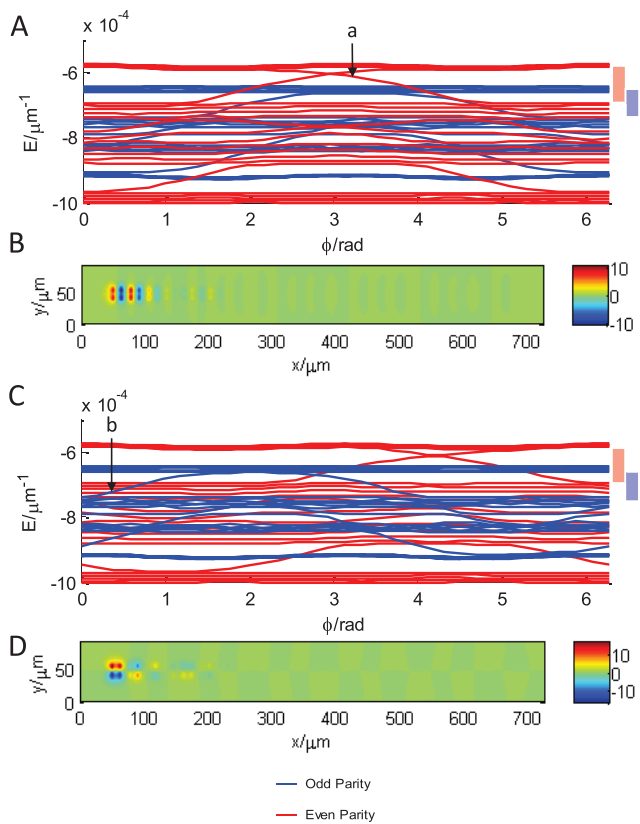}
\caption{(A) and (C) show energy spectra for our waveguide chain I and II, respectively. Boundary modes, depicted by ``$a$'' and ``$b$'' in (A) and (C), exist in the mini-gap and the corresponding wave functions, as shown in (B) and (D), are localized at the left boundary and possess even and odd parities respectively. The light red and light blue bar to the right indicate the range of interested mini-gap  of the respective parities mapped to Fig.\,\ref{fig_spectrum}B of the tight-binding model. A tight-binding model simulation using the same parameters are shown in the Appendix for comparison.}
\label{fig_eigenfunc}
\end{figure}

As shown in Fig.\,\ref{fig_model}B, four elliptical waveguides reside on the corners of a rectangle to simulate one unit cell in our tight-binding model. We place these unit cells one by one to form a chain, as described in the tight-binding model. The confining potential is provided by the deviation of the refraction index inside the waveguides from the background, $V=(\omega/c)\Delta n$, and the hopping term is determined by the overlap between two nearby sites, which can be controlled by adjusting the distance between them. Details of the relation between distance of different sites and energy level positions are discussed in the Appendix.

\begin{figure*}
\centering
\includegraphics[width=1\textwidth]{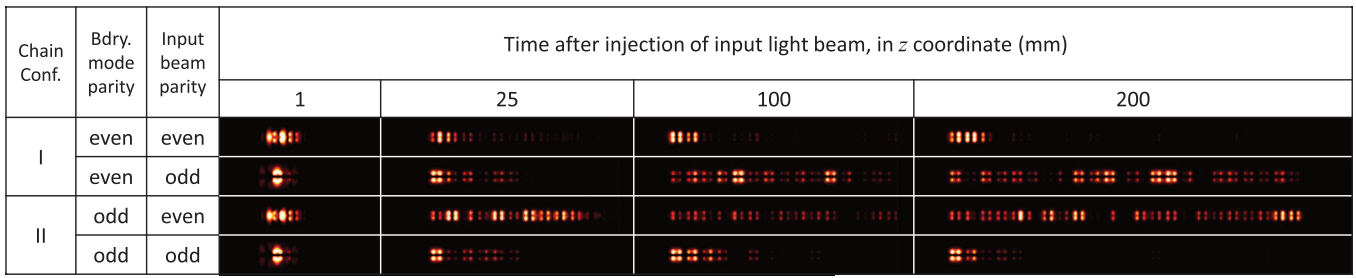}
\caption{ Numerical simulation of time-dependent evolution for the injection of the light beam $\psi^{\text{ini}}_{\pm}$ into the waveguide chains configuration I and II. The corresponding boundary mode parities and input beam parities are listed in the second and third column, respectively. Evolution time is expressed in unit of $z$-direction coordinate. Density profiles for wavefunctions at $z=$ 1, 25, 100 and 200 mm are presented. The dark parts in first three columns of the images are cropped.
}
\label{fig_anim}
\end{figure*}

For this waveguide chain with a finite number of unit cells ($M=23$), we first discuss our numerical simulation of the energy spectrum, which is shown in Fig.\,\ref{fig_eigenfunc}A and C for two waveguide chains with different parameter sets (See appendix for details of the choice of parameters). For simplicity, we label the waveguide chain for the energy spectrum in Fig.\,\ref{fig_eigenfunc}A and C as numbers I and II, respectively. Here we only focus on energy bands that correspond to $|\psi_\pm\rangle$ bands in our tight-binding model.  The energy spectrum of our continuum also reveals a sub-band structure with mini-band gaps for both parity bands and a detailed comparison between continuum model and tight-binding model is shown in the Appendix.  In short, we find that the continuum and tight-binding models strongly agree. For both the bands with even (red color) and odd (blue color) parities, they are split into three mini-bands, separated by mini-gaps.
Within the mini-gap, boudary modes can exist for both parity bands. Examples of boudary modes are marked by a for even parity bands in Fig.\,\ref{fig_eigenfunc}A (waveguide configuration I) and b for odd parity bands in Fig.\,\ref{fig_eigenfunc}C (waveguide configuration II), both at the left edge of the chain. For these two states, the corresponding wave functions are shown in Fig.\,\ref{fig_eigenfunc}B and D, respectively, from which one can clearly see that both modes are highly localized with a penetration length around twice of the lattice constant. We emphasize that we need to carefully choose the value of $\phi$ in Fig. \ref{fig_eigenfunc}A (Fig. \ref{fig_eigenfunc}C), and at that $\phi$ value, only one boudary mode ``$a$'' (``$b$'') at that edge is supported and this is important for the experiments of parity selected time evolution of edge modes described below.

%

We next inject a wave packet into one end of the waveguide chain and study time evolution of wave packets with different parities in this chain by solving the time-dependent Schr\"odinger equation, Eq.\,\ref{eq_schrodinger}, numerically. Technique details of our simulation are included in the appendix. The initial wave packet is chosen to be in Gaussian form, given by
\begin{equation}\label{eq_Gaussianbeam}
	\psi^{\text{ini}}_{\pm}(x,y)=\mathcal{N}\exp\left(-\frac{(x-x_0)^2}{2\sigma_x^{2}}-\frac{(y-y_0)^2}{2\sigma_y^{2}}\right) f_{\pm}(x,y)
\end{equation}
where $\mathcal{N}$ is a normalization factor, $(x_0,y_0)$ is the center of Gaussian form and chosen to be the center of the first two unit cells. The function $f$ describes the parity of the injected light beam and chosen to be $f_-(x,y)=e^{ik_x (x-x_0)}$ for even parity mode and $f_+(x,y)=\sin (k_y (y-y_0))$ for odd parity mode. The size of the envelope function is tuned to cover the majority of the mode wavefunction. We inject an even beam $\psi^{\text{ini}}_{-}(x,y)$ into the waveguide chain I with $\phi=1.15 \pi$ (the value of $\phi$ for the mode ``$a$'' in Fig. \ref{fig_eigenfunc}A), and the corresponding time evolution is shown in the first row of Fig.\,\ref{fig_anim}, from which one can see that the wave packet keeps localized after a long time evolution. In contrast, when we inject $\psi^{\text{ini}}_{+}(x,y)$ into the same waveguide configuration, the light spreads into the bulk of the waveguide chain, as shown in the second row of Fig.\,\ref{fig_anim}. For chain II with $\phi=0.25 \pi$ (the value of $\phi$ for the mode ``$b$'' in Fig. \ref{fig_eigenfunc}C), the light will get delocalized with initial wave packet $\psi^{\text{ini}}_{-}(x,y)$, but localized with initial wave packet $\psi^{\text{ini}}_{+}(x,y)$, as shown in the last two rows of Fig.\,\ref{fig_anim}. Therefore, we conclude that the localization of edge modes at the boundary sensitively depends on their mirror parities, and our numerical simulation also provides an approach to probe topological crystalline protection of boundary modes in realistic experiments.


\section{ Discussion and conclusion }
In conclusion, based on a tight-binding model and continuum numerical simulation, we propose an experimental setup of a photonic lattice to realize topological crystalline protection of boundary modes in photonic systems. Numerical simulation also suggests a possible experimental configuration to detect topological edge modes, of which the edge localization behavior depends on the parities of both the injected wave packets and localized edge modes. In addition, we would like to emphasize that two degenerate states realized in our system can also be regarded as ``pseudo-spin'' and thus provide a natural platform to construct SU(2) Landau levels \cite{li2013topological} and spin-orbit coupling in future studies.  The fact that the model was simulated with both tight-binding and more experimentally realistic continuum simulations suggests that can be implemented in a scheme similar to previously realized time-reversal-broken topological phases in photonic lattices.  This will open the door to the exploration of topological crystalline protection in interacting systems based on the nonlinear optical response of the ambient medium (giving rise to the nonlinear Schr\"odinger / Gross-Pitaevskii equation), and entangled quantum walks in such phases (based on injecting entangled photons into the structure).  A question of central importance in TCI physics is the question of disorder: since edge state protection is achieved using a symmetry that is easily broken (i.e., crystalline/mirror symmetry) by disorder, what is the nature of the robustness of these states?  Will all protection simply break down, or is it preserved in an ensemble-sense?  These are fundamental questions where photonics provides a unique and versatile path forwards.

\section*{ACKNOWLEDGMENTS}
We would like to acknowledge helpful discussions with Ling Lu, Marin Solja\v{c}i\'c, Rui-Xing Zhang, Jiabin Yu and Yang Ge. C.-X.L. acknowledges the support from Office of Naval Research (Grant No. N00014-15-1-2675).  M.C.R. acknowledges the support of the National Science Foundation under grant number ECCS-1509546.

\begin{appendix}


\section{Mapping to the 2D topological mirror insulator model}
In this section of the appendix, we will describe the detailed form of 2D topological mirror insulator model, which is mapped from the 1D topological mirror insulator model in the main text. The 2D tight-binding Hamiltonian is given by
\begin{eqnarray}
H &=& \sum_{xw,\alpha\beta} \left[\left( c^{\alpha\dag}_{x, w+1}  \hat {T}^{\alpha\beta}_x c^\beta_{x, w} + c^{\alpha\dag}_{x,w+1} \hat {T}^{\alpha\beta}_w c^{\beta}_{x,w} \right) + \text{h.c.}\right] \nonumber\\
&+& \sum_{xw,\alpha\beta } c^{\alpha\dag}_{x,w} \hat {U}_{\alpha\beta }  c^\beta_{x,w},
\end{eqnarray}
with
\begin{eqnarray}
& \hat T_x & = \left(
\begin{array}{cccc}
 0 & t & t' & 0 \\
 0 & 0 & 0 & 0 \\
 0 & 0 & 0 & 0 \\
 0 & t' & t & 0 \\
\end{array}
\right)
\nonumber \\
& \hat T_w & = \frac{\lambda}{4}
\left(
\begin{array}{cccc}
 \cos \theta & i \sin \theta & -\cos \theta & -i \sin \theta \\
 i \sin \theta & \cos \theta & -i \sin \theta & -\cos \theta \\
 -\cos \theta & -i \sin \theta & \cos \theta & i \sin \theta \\
 -i \sin \theta & -\cos \theta & i \sin \theta & \cos \theta \\
\end{array}
\right)\nonumber\\
&& +
\left(
\begin{array}{cccc}
 0 & \gamma_0 & 0 & \gamma_0 \\
 \gamma_0 & 0 & \gamma_0 & 0 \\
 0 & \gamma_0 & 0 & \gamma_0 \\
 \gamma_0 & 0 & \gamma_0 & 0 \\
\end{array}
\right)
\nonumber \\
& \hat U & = V_0 \cdot I_{4\times 4}. \nonumber
\end{eqnarray}
satisfying $2\gamma_0 >> \lambda$. Here $\gamma_0$ plays a similar role as that in the 1D model. $\gamma_0$ ensures the $\psi_t$ and $\psi_b$ states are away from the $\psi_\pm$ bands. Here $\alpha,\beta=1,\cdots,4$ are indices for four sites in one unit cell. $I$ is the identity matrix. We will show that $\theta$ is linked to a virtual external gauge field, and the parameters $t$ and $t'$ here match the definition in Eq.\,\ref{eq_Matrix_form_Hamiltonian}.

We can project the Hamiltonian onto the subspace of $d_{\pm}$ basis, which is written as
\begin{align}
\begin{split}
H &= \sum_{\pm,xw} \\
&\quad V_0 d^{\pm\dag}_{x,w} d^\pm_{x,w} + \left( \frac{-t'\pm t}{2} d^{\pm\dag}_{x+1,w}d^\pm_{x,w} + \frac{\lambda}{2} e^{\pm i\theta} d^{\pm\dag}_{x,w+1} d^\pm_{x,w} \right)\\
&\quad + \text{h.c.}\\
\end{split}
\end{align}
where $\pm$ is for odd and even subspaces, respectively. The parameter $\theta$ could be interpreted as the phase shift due to an external gauge field, where the field directions for the odd and even subspaces are opposite. We may choose the phase factor as $\pm \theta = \pm 2\pi b x$ for two parity subspaces, which correponds to the Landau gauge of a magnetic field in the $x-w$ plane in both the even and odd subspaces.

Substituting the Fourier transform along the $w$ direction $d^\pm_{x,w} = \sum_\phi e^{i \phi} d^\pm_{x,\phi}$ and $d^{\pm \dag}_{x,w} = \sum_\phi e^{-i \phi} d^\pm_{x,\phi}$, one can get

\begin{align}
\begin{split}
H &= \sum_{\phi} H(\phi) = \sum_{x,\phi,\pm} \\
&\quad V_0 d^{\pm\dag}_{x,\phi} d^\pm_{x,\phi} + \left(t_\pm d^{\pm\dag}_{x+1,\phi}d^\pm_{x,\phi} + \frac{\lambda}{2} e^{\pm i\theta} e^{i \phi} d^{\pm\dag}_{x,\phi} d^\pm_{x,\phi} \right) + \text{h.c.} \\
&= \sum_{x,\phi,\pm} \\
&\quad  t_\pm d^{\pm\dag}_{x+1,\phi}d^\pm_{x,\phi} + \text{h.c.} + \left( V_0 + \lambda \cos(\pm 2\pi b x + \phi)\right) d^{\pm\dag}_{x,\phi} d^\pm_{x,\phi}.
\end{split}
\end{align}
We find that $H(\phi)$ takes the same form of Eq.\,\ref{eq_AAM}. From the Fourier transform, the phase parameter $\phi$ in the original model plays a role of the momentum in the extra dimension $w$.

\section{Numerical Simulation}
In this section, we will describe the details of our numerical simulation of time dependent Schr\"odinger equation Eq.\,\ref{eq_schrodinger}, as well as the parameter setup of our waveguide chains I and II.
In the numerical simulation, the lattice constant is set to 28 $\mu$m. The separations between center of sites in the same unit cell is roughly 12 $\mu$m in $x$ and 13 $\mu$m in $y$ direction.  The other system parameters are the wavelength ($633$nm); the change in refractive index, $\Delta n = 7.2 \times 10^{-4}$; and $n_0=1.45$ (for silica). The system is discretized into a grid of $60 \times 120$ pixels for each unit cell on average.

The shape of the potential well for each site is given by
\begin{equation}\label{eq_cubic_potential_well}
V = -V_1\left(\exp\left(-\frac{x^2}{2\sigma_x^{'2}}-\frac{y^2}{2\sigma_y^{'2}}\right)\right)^3
\end{equation}
where $\sigma_x^{'}=1.9$ $\mu$m, $\sigma_y^{'}=5.5$ $\mu$m. The parameter $V_1=(\omega/c)\Delta n$ determines the depth of the potential well. Our numerical test shows that the on-site energy is linearly proportional to this depth within the energy range that we are interested in.

\begin{figure}
\centering
\includegraphics[width=0.5\textwidth]{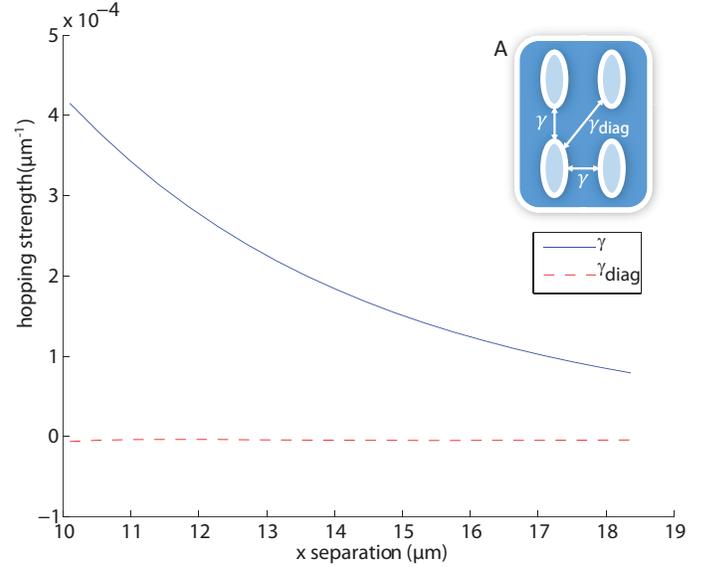}
\caption{Hopping strength $\gamma$ as a function of $x$ direction separation between two sites. The diagonal hopping $\gamma_\text{diag}$ strength is also calculated while keeping the hopping strength along $y$ direction to be $\gamma$ by modulating $y$ direction separation. The magnitude of $\gamma_\text{diag}$ is negligible. The inset (A) shows the two hopping terms inside a unit cell.}
\label{fig_x_gamma}
\end{figure}

The overlap between the wave functions for neighboring sites, which contributes to the hopping parameter, drops exponentially as a function of distance, as shown in Fig.\,\ref{fig_x_gamma}. From this numerical test, we find the hopping between neighboring sites $\gamma$ dominates over the hopping along the off-diagonal direction $\gamma_{\text{diag}}$ (shown in the inset of Fig.\,\ref{fig_x_gamma}), and thus we can neglect the off-diagonal hopping term between the sites 1 and 3 (2 and 4) and obtain the Hamiltonian Eq.\,\ref{eq_Matrix_form_Hamiltonian}. By controlling carefully the distance between different sites, four elliptical waveguides in one unit cell can indeed reproduce four eigen states with the desired energy levels as in our tight-binding model. By interpolation on Fig.\,\ref{fig_x_gamma}, we can establish the relation between the parameters described in the tight-binding model and the realistic parameters, such as the position of the potential well and refraction index in each site, in the continuum simulation model. Thus we are able to map the configuration in numerical simulation to the parameters in our tight-binding model. The parameters of the two models fall in the same range if we take the unit of hopping parameters in the tight-binding model to be mm$^{-1}$.

\begin{figure}
\centering
\includegraphics[width=0.5\textwidth]{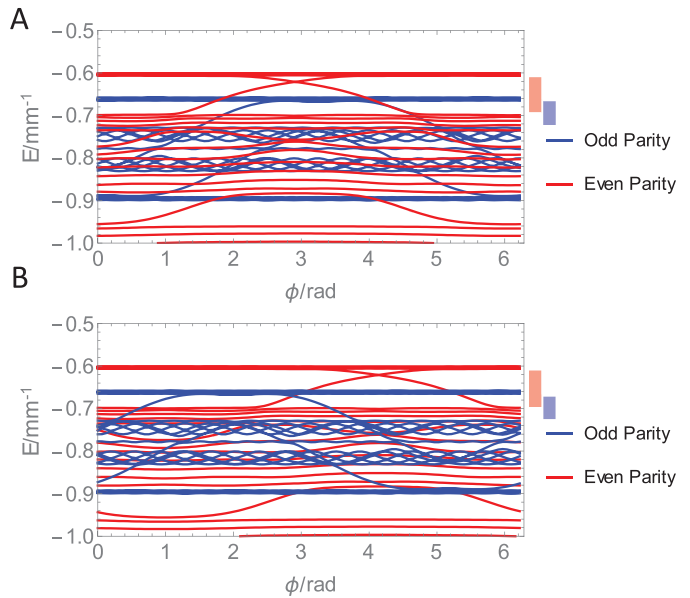}
\caption{Tight-binding simulation of chain I (A) and chain II (B). Compared with Fig.\,\ref{fig_eigenfunc}A and C, the positions of boundary modes appear to be the same. The light red and light blue bar to the right indicate the range of interested mini-gap of the respective parities, corresponding to the gap in Fig.\,\ref{fig_spectrum}B.}
\label{fig_tb_equivalence}
\end{figure}

If we focus on the structure of $\hat{V}_m$ in Eq.\,\ref{eq_Matrix_form_Hamiltonian} with the condition $\gamma_x = \gamma_y = \gamma_0$, the commutation relationship of $\left[\hat{V}_m(\phi), \hat{V}_m(\phi+\pi)\right] = 0$ suggests that there will be multiple boundary states within the band expanded by $\ket{\psi_+}$ and $\ket{\psi_-}$. For an arbitrary $\phi$, there could exist more than one boundary mode with different energies, as shown in Fig.\,\ref{fig_tb_equivalence}. A similar argument can be applied to numerical simulations and results in the excitation of multiple modes when the input beam is injected on the edge. To find a specific $\phi$ where there is only one boundary mode occupying a boundary, an integer $m_0$ setting the index of the leftmost unit cell is introduced into the tight-binding model and it does not change the topological properties of the system\cite{kraus2012topological}.
By carefully tuning $m_0$, we are able to find the configurations with only one boundary mode of one mirror parity.

\begin{table}
    \begin{tabular}{l|l}
    Parameter					& Value \\ \hline
    $V_0$/mm$^{-1}$				& -0.8\\
    $\lambda$/mm$^{-1}$			& 0.1\\
    $\gamma_{x0}$/mm$^{-1}$		& -0.275    \\
    $\gamma_{y0}$/mm$^{-1}$		& -0.3      \\
    $t$/mm$^{-1}$				& -0.13   \\
    $t'$/mm$^{-1}$				& -0.062  \\
    $b$							& $\sqrt{5}$ \\
    $M$							& 23\\
    $m_0$						& 22 for chain I, 0 for chain II \\
    \end{tabular}
    \caption{Parameters in tight-binding model to generate Fig.\,\ref{fig_tb_equivalence}. The values are fitted from the mapping between the tight-binding model and the continuum simulation.}
        \label{tb_tightbinding_params}
\end{table}

With the mapping of parameters between tight-binding model and continuum simulations, we can also identify the required wave-guide chain configurations for our continuum simulations. The dispersion spectra for the chains I and II have been given for continuum model in Fig. \ref{fig_eigenfunc}A and C. As a comparison, we also show the spectra of the chains I and II in tight-binding model in Fig.\,\ref{fig_tb_equivalence}. The positions of boundary modes for both parities match with the continuum simulations, although band widths and energies reveal minor difference. This is due to neglecting the long-range hopping and the correction of $t$ from the change of separations within nearby sites. Since we are only interested in boundary modes, this discrepancy is not important for our purpose. The detailed parameters used for the tight-binding model is listed in Table\,\ref{tb_tightbinding_params}.

The normalized input beam Eq.\,\ref{eq_Gaussianbeam} is treated as the initial wave function at time=0. For each timestep $dt$ in the simulation, the real-space wave function is multiplied by the time evolution operator $e^{-iHdt}$, where the potential part of Hamiltonian $H$ is discretized from the sum of the cubic-Gaussian potential well functions Eq.\,\ref{eq_cubic_potential_well} and the Fourier transform of the kinetic part. The reflection of light on the boundary of the grid is manually reduced to emulate experimental conditions. The timestep is set to 5um/$(c/n_0$).

\section{Topological invariant}
In this section, we will describe how to define mirror Chern number in our system, which follows the definition of Chern number for a quasicrystal system in Ref. \onlinecite{kraus2012topological}. For the quasicrystal system with one real dimension and one quasi-periodic term, one can define a periodic approximated Hamiltonian in a quasi-1D lattice with number of sites $M$ as
\begin{equation}\label{eq_Hamb}
	H_b(\phi,\xi)=\sum_{\alpha\beta,m} \left[ c_{m}^{\alpha\dag} \bar{ \hat{V}} ^{\alpha\beta}_{m} c^{\beta}_{m}+\left( c_{m}^{\alpha\dag}  \hat T^{\alpha\beta}_{m} e^{i\xi/M} c^{\beta}_{m+1}+\text{h.c.}\right) \right]
\end{equation}
where the lattice constant is set to 1. $\bar{ \hat{V_m}}$ is obtained by making the substitution in $ \hat{V_m}$ in Eq.\,\ref{eq_Matrix_form_Hamiltonian} by $b \rightarrow \bar{b}_M= \lfloor b\cdot M \rfloor /M$, where $\lfloor \cdot \rfloor$ is the floor function. $\bar{b}_M$ is essentially an rational approximation of the parameter $b$. Introducing $\xi$ implies a twisted boundary condition along $m$ direction.\cite{niu1987quantum} In the limit $\xi \rightarrow 0$ and $M \rightarrow \infty$, $H_b$ coincides with $H$ in Eq.\,\ref{eq_Ham0}. It is shown that for a finite large $M$ the topological properties of $H$ is the same as those of $H_b$.

Next, we follow the Ref. \onlinecite{kraus2012topological} to define topological invariants of this system through the projector operator for parity $p = +,-$, given by
\begin{equation}\label{eq_projector}
P_p(\phi,\xi) = \sum_{\substack{E_m<E_{\text{gap}}\\\text{parity}=p}} \ket{m}\bra{m}
\end{equation}
where $\ket{m}$ is the eigenstate of $H_b$ with energy $E_m$, and the summation is performed over all the bands with parity $p$ below the energy $E_{\text{gap}}$. $E_{\text{gap}}$ is the center energy of a gap. The Chern number for the bands with parity $p$ is defined by \cite{avron1983homotopy,kraus2012topological}
\begin{equation}\label{eq_ChernNumber}
\nu_p = \frac{1}{2\pi i} \int_0^{2\pi} C_p(\phi, \xi) \;d\phi d\xi
\end{equation}
where
\begin{equation}
C_p(\phi, \xi) = \text{Tr}\left(P_p \left[\frac{\partial P_p}{\partial \phi}, \frac{\partial P_p}{\partial \xi}\right]\right).
\end{equation}
The above expression is equivalent to the famious Thouless-Kohmoto-Nightingale-Nijs formula \cite{thouless1982quantized}
. With this definition of Chern number, the mirror Chern number can be defined as
\begin{equation}
\nu = \nu_+ - \nu_-,
\end{equation}
which characterizes topological mirror insulator phase for a system with mirror symmetry.

\end{appendix}



%

\end{document}